\documentclass[12pt,twoside]{article}

\usepackage{amsmath}

\input BoxedEPS

\SetRokickiEPSFSpecial  

\HideDisplacementBoxes

\numberwithin{equation}{section}

\makeatletter
\def\underbracket{%
    \@ifnextchar [ %
        {\@underbracket}%
        {\@underbracket [\@bracketheight]}%
}
\def\@underbracket[#1]{%
    \@ifnextchar [ %
        {\@under@bracket[#1]}%
         {\@under@bracket[#1][0.4em]}%
}
\def\@under@bracket[#1][#2]#3{
    \mathop {%
        \vtop {%
            \m@th \ialign {%
                ##\crcr $\hfil \displaystyle {#3}\hfil $%
                   \crcr \noalign %
                {\kern 3\p@ \nointerlineskip }%
                \upbracketfill {#1}{#2}
                   \crcr \noalign %
                {\kern 3\p@ }%
            }%
        }%
    }%
    \limits%
}
\def\upbracketfill#1#2{%
    $\m@th \setbox \z@ \hbox {$\braceld$}
    \edef\@bracketheight{\the\ht\z@}\bracketend{#1}{#2}
    \leaders \vrule \@height #1 \@depth \z@ \hfill
    \leaders \vrule \@height #1 \@depth \z@ \hfill%
    \bracketend{#1}{#2}$%
}
\def\bracketend#1#2{\vrule height #2 width #1\relax}
\makeatother

\makeatletter
\def\overbracket{\@ifnextchar [ {\@overbracket} {\@overbracket
[\@bracketheight]}}
\def\@overbracket[#1]{\@ifnextchar [ {\@over@bracket[#1]}
{\@over@bracket[#1][0.3em]}}
\def\@over@bracket[#1][#2]#3{
\mathop {\vbox {\m@th \ialign {##\crcr \noalign {\kern 3\p@
\nointerlineskip }\downbracketfill {#1}{#2}
                              \crcr \noalign {\kern 3\p@ }
                              \crcr  $\hfil \displaystyle {#3}\hfil $%
                              \crcr} }}\limits}
\def\downbracketfill#1#2{$\m@th \setbox \z@ \hbox {$\braceld$}
                  \edef\@bracketheight{\the\ht\z@}\downbracketend{#1}{#2}
                  \leaders \vrule \@height #1 \@depth \z@ \hfill
                  \leaders \vrule \@height #1 \@depth \z@ \hfill
\downbracketend{#1}{#2}$}
\def\downbracketend#1#2{\vrule depth #2 width #1\relax}
\makeatother

\newcommand{\ket}[1]{\left|#1\right\rangle}

\def\wt{\widetilde}

\begin{document}


\title{Gravitational scattering of massless scalars in QFT and superstring theory}
\author{J.~Pa\v sukonis\thanks{jurgis@mppmu.mpg.de}\\
\small Max-Planck-Institut f\"ur Physik\\
\small F\"ohringer Ring 6\\
\small 80805 M\"unchen\\[-0.25in]
}
\date{}
\maketitle

\pagestyle{myheadings}
\markboth{J. Pa\v sukonis}{Gravitational scattering of massless scalars in QFT and SST}
\thispagestyle{empty}

\begin{abstract}
\noindent
In this paper we perform the calculation of the gravitational scattering amplitude for 4 massless
scalars in quantum field theory and Type II superstring theory. We show that the results agree,
providing an example of how gravity is incorporated in the superstring theory. During the
calculation we quantize gravitational action to derive graviton propagator and interaction vertex
with massless scalar. We also calculate general 3-point and 4-point scattering amplitudes in SST
for open and closed massless strings in NS sector.
\end{abstract}

\section{Introduction}

The goal of this paper is to demonstrate an example of how gravity is incorporated in the superstring
theory (SST). In particular, we consider gravitational interaction between two massless scalar particles.
In quantum theory the quantity reflecting this interaction is the amplitude for a scattering process
of two scalar
particles by a graviton. First, we can perform this calculation in traditional quantum field theory.
Even though gravity can not be fully quantized in QFT, tree-level calculations can be done to give
results in agreement with classical theory. Second, we can use SST to get the amplitude
for the corresponding process. Here the particles corresponding to the massless scalars and gravitons
are massless closed strings of spin-0 and spin-2, respectively. With appropriate identification of
coupling constants the results of QFT and SST do agree, which confirms that SST includes gravitation.
In particular, it shows an example of how particles in SST having kinematic properties of a graviton
 (massless, spin-2)
also couple as gravitons.

This result is by no means original, and, actually, the argument of existance of gravity in SST can
be made much more general \cite{gs, pol2}. Also various results in this paper appear in other sources
and are cited. Therefore, this particular calculation is more of a way to go through some important
topics in QFT and SST, rather than an original work.

\section{Amplitude in QFT}

\begin{figure}[ht]
$$\BoxedEPSF{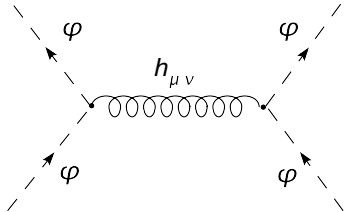 scaled 900}$$
\caption{two massless scalar scattering by a graviton.}
\label{fig:diag1}
\end{figure}

In this section we perform the calculation of the gravitational scattering of two massless scalars
in quantum field theory. We only consider the first-order tree-level process
$\phi\phi\rightarrow\phi\phi$, that is, with only one intermediate graviton. One Feynman diagram
representing such interaction is
Fig.~\ref{fig:diag1} (we notate a scalar particle with $\phi$ and graviton with $h$).

In order to calculate such amplitude we need two results from QFT: graviton propagator and 2-scalar/graviton
interaction vertex. We proceed now with deriving those values from the classical actions.

\subsection{Graviton propagator}

The gravitational action is \cite{gr,gs}:
\begin{equation}
S_{G}=\frac{1}{16\pi G}\int d^D x\, \sqrt{-g} R=\frac{1}{2 \kappa^2} \int d^D x\, {\cal L}_G,
\label{eq:sg}
\end{equation}
with
\begin{equation}
{\cal L}_G\equiv \sqrt{-g} R,\quad \kappa\equiv\sqrt{2\pi G}.
\label{eq:lg}
\end{equation}
Here $G$ is Newton's constant, $g$ is the determinant of $g_{\mu\nu}$ and $R$ is the Ricci scalar.
 This particular normalization of $S_G$ is needed to get the correct Einstein's equation when
we add conventionally normalized matter action. Note that we are working in arbitrary number of
dimensions $D$.

To be able to do perturbative calculation (to first order in our case) we assume the gravitational
field is weak, that is, the metric is almost flat \cite{gr}:
\begin{equation}
g_{\mu\nu}=\eta_{\mu\nu}+\beta h_{\mu\nu},\quad |\beta h_{\mu\nu}|\ll 1,
\label{eq:gnh}
\end{equation}
note that we use the convention for $\eta_{\mu\nu}$, that differs from \cite{ps} but agrees with
the other references listed:
\begin{equation}
\eta_{\mu\nu}=\text{diag}(-,+,+,+,...).
\label{eq:eta}
\end{equation}
By the arbitrary normalization constant $\beta$ in (\ref{eq:gnh}) we can redefine the normalization of $h$ - we will
need this to get the correctly normalized graviton states in quantum field theory. In
non-quantum general relativity it is usually taken $\beta=1$ in this expansion.
Note that $h_{\mu\nu}$, like $g_{\mu\nu}$, is symmetric. We also need an expression for the
corresponding inverse-metric:
\begin{equation}
g^{\mu\nu}=\eta^{\mu\nu}-\beta h^{\mu\nu}+O(h^2),
\end{equation}
where indices on $h_{\mu\nu}$ are raised with $\eta^{\mu\nu}$.

Now we Taylor-expand ${\cal L}_G$ in
(\ref{eq:lg}) to the second power in $h$. Up to a total derrivative the result is (this differs
by a sign from that in \cite{gr})
\begin{equation}
{\cal L}_{h}=\frac{\beta^2}{2}\left[
  h \partial_\mu\partial_\nu h^{\mu\nu}-
  h^{\mu\rho}\partial_\mu\partial_\nu {h^\nu}_\rho+
  \frac{1}{2}h_{\mu\nu}\partial^2 h^{\mu\nu}-
  \frac{1}{2}h\partial^2 h
\right].
\label{eq:lg2}
\end{equation}
Here $h={h^\mu}_\mu$ and
$\partial^2=\partial_\mu\partial^\mu$. The total derrivative in ${\cal L}_G$ is irrelevant since
it gets integrated in (\ref{eq:sg}) over all space and we assume the boundary terms to vanish.
This quadratic Lagrangian ${\cal L}_h$ in (\ref{eq:lg2}), when quantized, describes a massless
particle,
namely, the graviton\footnote{The problem with quantum gravity only appears when we consider higher
orders in ${\cal L}_G$ expansion and loop Feynman diagrams - the theory turns out to be
non-renormalizable}.

Now we consider quantizing this gravitational action, but let's fix the $\beta$ in (\ref{eq:gnh}) before
we go further. It turns out that the value that gives correctly normalized states of $h$ is \cite{gs}:
\begin{equation}
\beta=2\kappa.
\end{equation}
With this $\beta$, the $\kappa$'s cancel in
$S_G$ and we get:
\begin{equation}
S_h\equiv\int d^D x\,
\left[
  h \partial_\mu\partial_\nu h^{\mu\nu}-
  h^{\mu\rho}\partial_\mu\partial_\nu {h^\nu}_\rho+
  \frac{1}{2}h_{\mu\nu}\partial^2 h^{\mu\nu}-
  \frac{1}{2}h\partial^2 h
\right],
\end{equation}
which is the action that we will quantize.

We proceed with the quantization of $h$ field using Feynman path integral approach
(this part, especially Faddeev-Popov gauge fixing, is based on Chapter 9 in \cite{ps}).
We need to consider the
quantity
\begin{equation}
Z=\int {\cal D}h\, e^{i S_h[h]},
\label{eq:hZ}
\end{equation}
where ${\cal D}h$ integrates over all inequivalent field configurations. Once we have such
an expression, we can deduce the propagator for the field directly from the $S$ appearing in
(\ref{eq:hZ}).

The problem with $S_h$ is that there is a gauge transformation:
\begin{equation}
h^\xi_{\mu\nu}=h_{\mu\nu}+\partial_\mu \xi_\nu+\partial_\nu \xi_\mu,
\label{eq:hxi}
\end{equation}
for any field $\xi_\mu$, that leaves $S_h$ invariant.
Therefore, simple ${\cal D}h$ in (\ref{eq:hZ}) includes many
equivalent field configurations. To get the correct $Z$ we need to factor out from
(\ref{eq:hZ}) the integral over gauge transformations, leaving only
the integral over physically different configurations. This is done by Faddeev-Popov gauge
fixing procedure, which we briefly describe here.

First, we choose a gauge-fixing function $G(h)$
\footnote{The discussion here also works
for $G_\lambda(h)$ with an index, that is, for a collection of gauge-fixing functions.},
such that the constraint $G(h)=0$ would
pick out \emph{one} field configuration from each set of equivalent ones, that is, fix the gauge.
Then we can write (\ref{eq:hZ}) as \cite{ps}:
\begin{align}
Z=\int {\cal D}h\, e^{i S_h[h]}&=
\Delta
\int {\cal D}\xi\, \int {\cal D}h\, e^{i S_h[h]} \delta(G(h)),
\label{eq:zgauge1}\\
\Delta&\equiv\det \left( \frac{\delta G(h^\xi)}{\delta \xi} \right).
\end{align}
Here $h^\xi$ as in (\ref{eq:hxi}) and $\Delta$ is a functional determinant arising as
the Jacobian for the relevant change of variables. We must choose $G$ such that the $\Delta$
is independent of $h$, thus, just a constant. We see that we did factor out $\int {\cal D}\xi$,
the integral over gauge transformations, and the inner integral is constrained
by functional $\delta$, allowing only gauge-fixed, thus inequivalent, field configurations.

Now we consider a particular gauge-fixing function:
\begin{equation}
G_\lambda=\partial_\mu {h^\mu}_\lambda-\frac{1}{2}\partial_\lambda h - w_\lambda,
\end{equation}
where $w_\lambda$ is any field. Then $G_\lambda=0$ is a generalized Lorentz gauge condition
(it is Lorentz gauge for $w_\lambda=0$). We can not, though, use (\ref{eq:zgauge1}) for
direct calculations, because it is, due to the $\delta$-function, not in the standard form
(\ref{eq:hZ}). We fix that
as follows.  Since (\ref{eq:zgauge1}) is true for any $G$, we can
integrate the equation with $G_\lambda$ over $w$ with any normalized-function weighting and cancel the
$\delta$-function with this integral. In particular, we use:
\begin{equation}
N(\alpha) \int {\cal D}w\, \exp \left(i \int d^Dx\, \alpha\, w_\lambda w^\lambda \right)=1,
\end{equation}
with some real non-zero coefficient $\alpha$ and normalization
constant $N(\alpha)$ to integrate (\ref{eq:zgauge1}). Note that
this works on RHS because $\Delta$ is $w$-independent. We get:
\begin{align}
Z &= N \int {\cal D}w\, \exp \left(i \int d^Dx\, \alpha\, w_\lambda w^\lambda \right)
\Delta \int {\cal D}\xi\, \int {\cal D}h\, e^{i S_h[h]} \delta(F_\lambda(h)-w_\lambda) \nonumber\\
 &= N\Delta \left(\int {\cal D}\xi\right)
   \int {\cal D}h\, \exp \left(i S_h[h] + i \int d^Dx\, \alpha\, F_\lambda F^\lambda \right),
\end{align}
where
\begin{equation}
F_\lambda(h)\equiv \partial_\mu {h^\mu}_\lambda-\frac{1}{2}\partial_\lambda h.
\end{equation}
Thus in effect the result of Faddeev-Popov procedure is adding a new term
to the Lagrangian and factoring out an overall (infinite) constant from Z:
\begin{align}
Z&=C   \int {\cal D}h\, \exp \left(i S'_h \right),\\
S'_h&=\int d^Dx\, \left(2{\cal L}_h+\alpha\, F_\lambda F^\lambda\right).
\label{eq:shg1}
\end{align}
All the factors, including gauge integral, are now contained in an overall factor $C$,
which is irrelevant for
calculating expectation values. What we are left with is the standard form of $Z$ with
action $S'_h$, from which we can correctly deduce the graviton propagator.

The calculations must work for any value of $\alpha$, but we can set it to simplify our
Lagrangian (\ref{eq:lg2}). The additional term, up to a total derrivative, is:
\begin{equation}
\alpha\, F_\lambda F^\lambda=
\alpha \left(h \partial_\mu\partial_\nu h^{\mu\nu}-
  h^{\mu\rho}\partial_\mu\partial_\nu {h^\nu}_\rho-
  \frac{1}{4}h\partial^2 h\right).
\end{equation}
Choosing $\alpha=-1$, action (\ref{eq:shg1}) becomes:
\begin{equation}
S'_h=\frac{1}{2}\int d^Dx\,
\left(h_{\mu\nu}\partial^2 h^{\mu\nu}-\frac{1}{4}h\partial^2 h\right),
\label{eq:shg2}
\end{equation}
which we can finally use to get the graviton propagator.

In general, we can deduce the propagator from an action as follows.
Suppose we have a real field $\phi$ described by:
\begin{equation}
S=\frac{1}{2} \int d^D x\, \phi(x)\, Q\, \phi(x),
\end{equation}
with some differential operator $Q$. In case $\phi$ has indices, $\phi Q\phi$ is matrix
multiplication. The propagator $D(x-y)=\langle\phi(x)\phi(y)\rangle$ then satisfies:
\begin{equation}
Q\, D(x-y) = i \delta(x-y) {\bf I},
\label{eq:qdx}
\end{equation}
where ${\bf I}$ is matrix identity. With $Q$ and $D$ transformed into momentum space
(\ref{eq:qdx}) becomes:
\begin{equation}
\wt{Q}(k)\wt{D}(k)=i {\bf I}.
\label{eq:qdi}
\end{equation}

Now we apply these identities to $S'_h$. We can rewrite (\ref{eq:shg2}) as:
\begin{align}
S'_h&=\frac{1}{2}\int d^Dx\,
h_{\mu\nu} Q^{\mu\nu;\rho\sigma} h_{\rho\sigma}, \\
Q^{\mu\nu;\rho\sigma}&=\frac{1}{2}\left(
\eta^{\mu\rho}\eta^{\nu\sigma}+\eta^{\mu\sigma}\eta^{\nu\rho}-\eta^{\mu\nu}\eta^{\rho\sigma}
\right)\partial^2,
\end{align}
where we have explicitly symmetrized $Q$ in $\{\mu\nu\}$ and $\{\rho\sigma\}$, since $h_{\mu\nu}$
is always symmetric, thus antisymmetric matrix component acting on it is irrelevant. In
momentum space this looks as
\begin{equation}
\wt{Q}^{\mu\nu;\rho\sigma}=-\frac{k^2}{2}\left(
\eta^{\mu\rho}\eta^{\nu\sigma}+\eta^{\mu\sigma}\eta^{\nu\rho}-\eta^{\mu\nu}\eta^{\rho\sigma}
\right),
\end{equation}
and the identity in (\ref{eq:qdi}), since we are working with symmetric matrices, is:
\begin{equation}
{\bf I}^{\mu\nu}_{\rho\sigma}=
\frac{1}{2}\left(\delta^\mu_\rho \delta^\nu_\sigma+\delta^\mu_\sigma \delta^\nu_\rho\right).
\end{equation}
Now we can find $\wt{D}$ satisfying:
\begin{equation}
\wt{Q}^{\mu\nu;\alpha\beta}\wt{D}_{\alpha\beta;\rho\sigma}=i {\bf I}^{\mu\nu}_{\rho\sigma},
\end{equation}
to be:
\begin{equation}
\wt{D}^{(h)}_{\mu\nu;\rho\sigma}(k)=-\frac{1}{2}
\left(\frac{i}{k^2-i\epsilon}\right)
\left(
\eta_{\mu\rho}\eta_{\nu\sigma}+\eta_{\mu\sigma}\eta_{\nu\rho}
-\frac{2}{D-2}\eta_{\mu\nu}\eta_{\rho\sigma}\right).
\label{eq:propagator}
\end{equation}
in agreement with \cite{bern}.
This is the final expression for momentum-space graviton propagator between polarizations
$(\mu \nu )$ and $(\rho \sigma )$. Term $-i \epsilon $ was added in the denominator as usual
to give the right behavior of the integral to position space.

\subsection{Scalar-graviton interaction}
Now we proceed with calculating the second component needed for our scattering calculation:
the scalar-graviton interaction vertex. Again we start with classical action and quantize
it.

First consider a general matter action added to the pure gravitational one (\ref{eq:sg}):
\begin{align}
S&=S_G+S_M,\\
S_M&=\int d^D x\, \sqrt{-g} {\cal L}_M.
\end{align}
This action includes matter-gravity interaction, because it has both metric and matter field
terms. Note that the Lagrangian ${\cal L}_M$ itself can contain metric
terms.

Again we want to consider weak gravitational field behavior, for which we expand $S_M$ in Taylor
series around the flat matric. As in (\ref{eq:gnh}), variation in metric is $\beta h_{\mu\nu}$, and
we will set $\beta=2\kappa$ when quantizing. We are interested only in first-order
interaction, which corresponds to vertices with only one graviton involved, so we expand $S_M$
up to the first order in $h$:
\begin{align}
S_M=(S_M)_{g=\eta}+\int d^D x\, (-\beta h^{\mu\nu})
\left(\frac{\delta S_M}{\delta g^{\mu\nu}}\right)_{g=\eta}+O(h^2).
\label{eq:232}
\end{align}
The first term describes non-interacting (gravitationally) matter:
\begin{equation}
S_{m}\equiv(S_M)_{g=\eta}=\int d^D x\, ({\cal L}_M)_{g=\eta}.
\end{equation}
The second describes first-order gravitational interaction and it can be expressed in terms
of a familiar quantity - energy-momentum tensor, defined as (\cite{gr}, up to a factor):
\begin{equation}
T_{\mu\nu}\equiv -\frac{2}{\sqrt{-g}}\frac{\delta S_M}{\delta g^{\mu\nu}}=
-2\frac{\partial {\cal L}_M}{\partial g^{\mu\nu}}+g_{\mu\nu}{\cal L}_M.
\label{eq:t}
\end{equation}
The interaction term in (\ref{eq:232}) is then:
\begin{equation}
S_I=\frac{\beta}{2}\int d^D x\, h^{\mu\nu}(T_{\mu\nu})_{g=\eta},
\label{eq:sit}
\end{equation}
and the total matter action is:
\begin{equation}
S_M=S_m+S_I+O(h^2).
\end{equation}

Now we go to our case of interest, the massless scalar field $\phi$. Such field
is described by the action:
\begin{align}
S_M&=-\frac{1}{2}\int d^D x\, \sqrt{-g}(g^{\mu\nu}\partial_\mu \phi\, \partial_\nu \phi),
\label{eq:sM}\\
{\cal L}_M&=-\frac{1}{2}g^{\mu\nu}\partial_\mu \phi\, \partial_\nu \phi.
\end{align}
We get the free field action $S_m$ by just replacing $g^{\mu\nu}$ by $\eta^{\mu\nu}$
in (\ref{eq:sM}), from which we could easily get the scalar field propagator.
For our calculation that is not necessary, though, so we concentrate on $S_I$, which
contains the interaction vertex we need. To use (\ref{eq:sit}) first we get calculate
the momentum-energy tensor for $\phi$ by (\ref{eq:t}):
\begin{equation}
T_{\mu\nu}=\partial_\mu\phi\partial_\nu\phi-
\frac{1}{2}g_{\mu\nu}(\partial_\lambda\phi\, \partial^\lambda\phi).
\end{equation}
Now plugging this $T_{\mu\nu}$ in (\ref{eq:sit}) with $\beta=2\kappa$ we have:
\begin{equation}
S_I=\kappa\int d^D x\, \left(h_{\mu\nu}\partial^\mu\phi\,\partial^\nu\phi-
\frac{1}{2}h_{\mu\nu}\eta^{\mu\nu}\partial_\lambda\phi\,\partial^\lambda\phi\right),
\end{equation}
the action which we use for quantization.

\begin{figure}[ht]
$$\BoxedEPSF{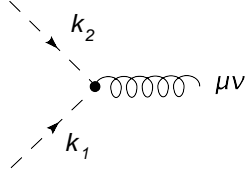 scaled 900}$$
\caption{$\phi(k_1)\phi(k_2)h_{\mu\nu}$ vertex}
\label{fig:vertex}
\end{figure}

In the Feynman path integral quantization the vertex values can be easily read off
directly from the action - they are basically just the coefficients of the corresponding
field products. In this $S_I$ we have a product of two $\phi$ fields and an $h_{\mu\nu}$
field, which does correspond to the vertex we want (Fig.~\ref{fig:vertex}).
We arbitrarily
chose the momenta to be incoming - the outgoing momentum is then represented by
negative-incoming. To get the vertex value in the momentum space we first substitute
$\partial_\mu\rightarrow ik_\mu$ (for an incoming momentum), with $k_\mu$ of the field the
derrivative acts on, which gives the term in $S_I$ as:
$$
-\kappa\left(k_1^\mu k_2^\nu-\frac{1}{2}\eta^{\mu\nu}k_1\cdot k_2\right)
\phi(k_1)\phi(k_2)h_{\mu\nu}(k_3).
$$
The value of the vertex is then given by this coefficient in front of the fields multiplied
by a factor of 2 from permuting identical lines $\phi$ and
an a factor of $i$ that we have in any vertex (it comes from $\exp(iS)$). Thus the vertex
amplitude is:
\begin{equation}
V_{(\phi\phi h)}^{\mu\nu}(k_1,k_2)=-i\kappa\left(k_1^\mu k_2^\nu+k_1^\nu k_2^\mu-
\eta^{\mu\nu}(k_1\cdot k_2)\right),
\label{eq:vertex}
\end{equation}
where we have again explicitly symmetrized the tensor multiplying $h_{\mu\nu}$.

\subsection{QFT scattering amplitude}
\begin{figure}[ht]
$$\BoxedEPSF{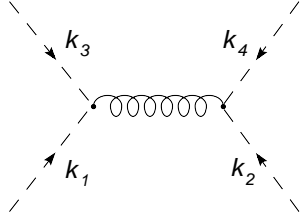 scaled 900}$$
\caption{Feynman diagram for the scattering process ($t$-channel)}
\label{fig:diag2}
\end{figure}

Now we have all the pieces to calculate the scattering amplitude (S-matrix) for our
process. Consider the Feynman diagram Fig.~\ref{fig:diag2}.
At first we perform the amplitude calculation only for this one ``channel" - the diagram
where $k_1$ and $k_3$ are connected at one vertex ($t$-channel), and then we construct
the full amplitude (which will include terms from two more channels) using this result.
Note that we marked all
$k_i$ as incoming - this is partly to make the comparison with the string theory
result easier. In applying our result based on this diagram to an actual process with initial
momenta $(k_1,k_2)$ and final momenta $(k'_1,k'_2)$ we would have to take $k_3=-k'_1$
and $k_4=-k'_2$ to reverse the direction.

The S-matrix for the diagram Fig.~\ref{fig:diag2} is (Chapter 4.5 in \cite{ps}):
\begin{align}
S_t&=(2\pi)^D\delta^D(\Sigma_i k_i)i{\cal M}_t \label{eq:s-matrix},\\
i{\cal M}_t&=V^{\mu\nu}_{(\phi\phi h)}(k_1,k_2)\wt{D}^{(h)}_{\mu\nu;\rho\sigma}(k_1+k_2)
V^{\rho\sigma}_{(\phi\phi h)}(k_3,k_4).
\label{eq:smatrix}
\end{align}
We can evaluate $i{\cal M}_t$ by plugging in values from (\ref{eq:propagator}) and
(\ref{eq:vertex}). Note that for our case of interest:
\begin{align}
\label{eq:2.44}
k_1^2=k_2^2=k_3^2=k_4^2=0,\\
k_1+k_2+k_3+k_4=0,
\end{align}
due to masslessness and momentum conservation ($\delta$-function in (\ref{eq:s-matrix}))
respectively. It is useful then to define ``Mandelstam variables" \cite{ps,pol1,zwiebach} :
\begin{align}
\label{eq:2.46}
s&\equiv -(k_1+k_2)^2=-2k_1\cdot k_2= -(k_3+k_4)^2=-2k_3\cdot k_4,\\
t&\equiv -(k_1+k_3)^2=-2k_1\cdot k_3= -(k_2+k_4)^2=-2k_2\cdot k_4,\\
u&\equiv -(k_1+k_4)^2=-2k_1\cdot k_4= -(k_2+k_3)^2=-2k_2\cdot k_3,
\end{align}
that also satisfy
\begin{equation}
s+t+u=0.
\end{equation}
Evaluating $i{\cal M}_t$ gives our desired amplitude for $t$-channel, which has a very
simple expression in terms of Mandelstam variables:
\begin{align}
i{\cal M}_t&=i\kappa^2 \frac{su}{t},\\
S_t&=i\kappa^2(2\pi)^D\delta^D(\Sigma_i k_i)\frac{su}{t}.
\end{align}
 Note that the there is a pole in $S$ at $t=0$, which should have been expected, since
$t=-(k_1+k_3)^2=-q^2=m^2$ of the
virtual graviton, and the pole appears when the virtual particle is on-shell, in this case,
$m^2=0$. It is because of the pole in $t$, which comes from the fact that $k_1$ and $k_2$ are connected
at a vertex, that this configuration is called $t$-channel.

\begin{figure}[ht]
$$\BoxedEPSF{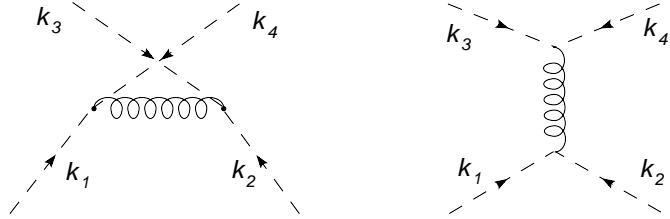 scaled 900}$$
\caption{Feynman diagrams for $u$-channel (left) and $s$-channel (right).}
\label{fig:diag3}
\end{figure}

As mentioned before, the amplitude $S_t$ is not the full amplitude because there are two
more Feynman diagrams contributing to the same observed scattering process
$\phi\phi\rightarrow\phi\phi$ mediated by a graviton. We get these by simply
permuting $k_i$'s in the diagram of Fig.~\ref{fig:diag2}
- the resulting diagrams are shown in Fig.~\ref{fig:diag3}.
The amplitudes are as in (\ref{eq:smatrix}), but with $k_i$ permuted, which causes ${s,u,t}$
permutation:
\begin{align}
i{\cal M}_u &=i\kappa^2 \frac{st}{u},\\
i{\cal M}_s &=i\kappa^2 \frac{tu}{s}.
\end{align}
These processes have poles in $u$ and $s$ (from $(k_1,k_4)$ and $(k_1,k_3)$ connections),
 therefore are called $u$-channel and $s$-channel.

The full scattering amplitude for the process is thus the sum of the three diagrams,
All the particles
involved are bosons, so there are also no relative minus signs from permutations, therefore, the
total amplitude is:
\begin{equation}
S=S_t+S_u+S_s=i\kappa^2(2\pi)^D\delta^D(\Sigma_i k_i)
\left(\frac{tu}{s}+\frac{us}{t}+\frac{st}{u}\right),
\label{eq:sqft}
\end{equation}
which is our final QFT result that we will compare to the string theory calculation.

\section{Amplitude in SST}

Now we proceed with the calculation of the scattering amplitude in superstring theory.
This calculation is strongly based on refs. \cite{pol1,pol2} and
many of the results in this section appear there.

We consider Type I open and Type II closed superstring theories, which are among the most
realistic string theories, and include our particles of interest - massless scalars
and massless spin-2 particles (gravitons). For our purposes heterotic superstring theory
could be substituted
for Type II, as it contains all the same properties we are interested in, and the
result for this scattering in comparison with QFT would be the same.

The superstring theories require a 10-dimensional spacetime. We take it to be flat, that
is, described by metric $\eta_{\mu\nu}$ as in (\ref{eq:eta}),
with 9 extended spatial dimensions. This does not represent a realistic model for our
universe, but
considering compactifications of 6 dimensions adds too many complications. We can still
compare our result in 10 dimensions with the QFT result, as we derived it for arbitrary
$D$. We will continue denoting the number of dimensions as $D$ in this section, but it should
be kept in mind that the results only make sense for
\begin{equation}
D=10.
\end{equation}

\subsection{General amplitudes in SST}
Here we will discuss qualitatively how the strings are described and, consequently, how
the amplitudes are calculated in string theory. The quantitative results needed for
the calculation, the string vertex operators and related field expectation values,
will not be derived, and taken from \cite{pol1,pol2} in the following
sections.

Strings moving in spacetime are described by a collection of \emph{fields} living on a
two-dimensional surface called \emph{world sheet}. To see this consider a motion of an
open classical string, which can be described by its position in spacetime
$X^\mu(\sigma,\tau)$, with \emph{two} parameters: $\sigma\in(0,a)$ parametrizes position
along the string and
$\tau\in(-\infty,\infty)$ is some parametrization of the motion in time. Then we can
interpret this parameter space as a two-dimensional space (world sheet), on which we have
$D$ scalar fields $X^\mu$. In the case of the open string, the world sheet is a strip
of finite width and infinite length, while for a closed string it would be an
infinitely long cylinder because of the identification $\sigma\sim\sigma+a$.

The quantization of a string is then just the quantization of the fields with the
appropriate action on the world sheet, which results in a two-dimensional QFT. In the
quantization, however, more fields than just $X^\mu$ are introduced: for a superstring we have
in addition $D$ anticommuting fields $\psi^\mu$, also the ``world-sheet metric" that
appears in the action is itself made into a dynamic field and, finally, we get
gauge symmetries which result in ghost fields while being fixed. Each of these fields
makes up an independent QFT, with the expectation values that we will cite when needed.

One important feature of the string action is a \emph{conformal symmetry} of the
world-sheet, which means that the world-sheets related by a conformal transformation are equivalent,
that is, they describe the same string process. This symmetry has many important consequences
and the QFT's with such symmetry are called conformal field theories (CFT's).
The one property of conformal symmetry relevant to our qualitative discussion is that we
essentially need to consider only different \emph{topologies} of the world sheets,
ignoring the exact shape\footnote{Conformal equivalence of world-sheets
is a bit narrower definition than topological equivalence (meaning that one can be deformed continuously
 into another). The difference, however, is ``small" in a sense that the parameter space
of conformally-inequivalent world-sheets within a certain topology is finite-dimensional.}.

\begin{figure}[ht]
$$\BoxedEPSF{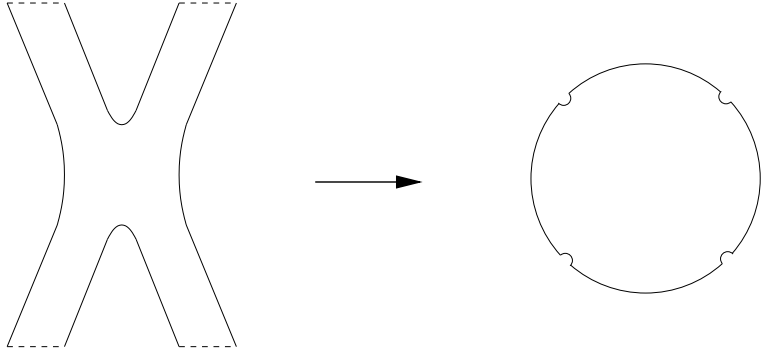 scaled 600}$$
\caption{World sheet of 4 open string interaction - a disk.}
\label{fig:4open}
\end{figure}

\begin{figure}[ht]
$$\BoxedEPSF{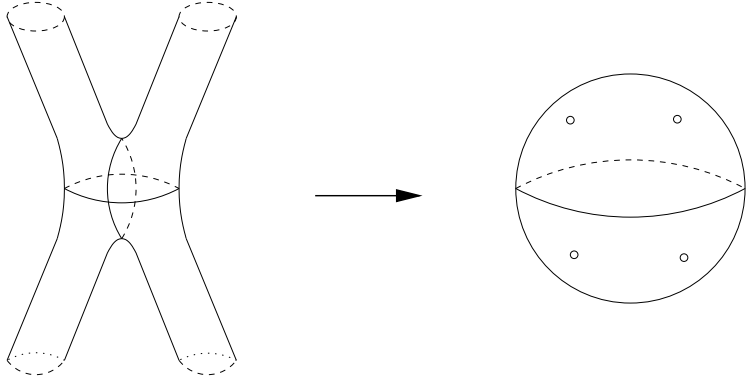 scaled 600}$$
\caption{World sheet of 4 closed string interaction - a spehere.}
\label{fig:4closed}
\end{figure}

Now we consider string interactions. The nice property of string theory is that interactions
are described in a natural way
by the same acion, only with considering non-trivial shapes of the world-sheets that
the fields live on. For example in Fig.~\ref{fig:4open} we see a world sheet for 4 open
strings interacting by an open string and in Fig.~\ref{fig:4closed} for 4 closed strings
interacting by another closed string. We can conclude that it is that particular
kind of intermediate string by looking at a shape of a cut through the world sheet in
the ``interaction region".


Shown with arrows are conformally equivalent world sheets
(it is easy to see that they are topologically equivalent).
In case of 4 open strings we get
a disk with 4 points missing at the boundary - these arise from 4 ``legs" extending
infinitely in time-like direction of the world sheet, where the boundary is missing. Similarly
for 4 closed strings the conformally equivalent surface is a sphere with 4 points missing.

Consider, finally, the calculation of the amplitude itself. By the method of Feynman path
integral quantization, it is given by the sum of $e^{i S}$
over all inequivalent configurations of world sheets ${\cal W}$ and fields
on them $\Phi$, with some specified initial and final states:
\begin{equation}
S=\int{\cal D}[{\cal W}]{\cal D}[\Phi]_{\Phi_0(z_0)}
\exp(i S[{\cal W},\Phi]).
\end{equation}
We denoted by $\Phi_0(z_0)$ the constraint on $\Phi$ that it has to satisfy at the set of points
$z_0$ (${\cal W}$-dependent) in the boundary of the world-sheet, where the initial and final
string states are defined. We can
always conformally transform the world-sheet so that, as in our examples, the initial and
final state strings appear as missing \emph{points} $z_i$, around which $\Phi_0$ has to be
defined. It
can be shown \cite{pol1} that equivalently we can substitute the initial and final state constraints
by a string \emph{vertex operators} ${\cal V}_i(z_i)$ at those points, and let $\Phi$ take
any values:
\begin{equation}
S=\int{\cal D}[{\cal W}]{\cal D}[\Phi]
\exp(i S[{\cal W},\Phi])
\prod_i {\cal V}_i(z_i).
\label{eq:sworlds}
\end{equation}
This says that the S-matrix amplitudes are the expectation values of vertex operators, which
we can think of as operators creating an incoming or outgoing string somewhere on the world sheet.

Consider now splitting the integral ${\cal D}[{\cal W}]$ into a sum of integrals within each
topology:
\begin{equation}
S=\sum_{\text{topologies}} \int{\cal D}[{\cal M}]{\cal D}[\Phi]
\exp(i S[{\cal W},\Phi])
\prod_i {\cal V}_i(z_i).
\label{eq:stopologies}
\end{equation}
The ${\cal M}$ is the so-called \emph{moduli} - the parameters within a topology distinguishing
conformally-inequivalent world sheets (this is where the small difference between topological
and conformal equivalence goes). We concentrate instead on the sum over topologies now.
The world-sheet topology tells us how many and what kind of strings (closed, open) are involved
in the process - it is the equivalent of Feynman diagrams in QFT. More complicated topologies
(that is, with more holes) will describe processes with more interactions and more intermediate
particles. This way we organized the calculation as perturbation series, and we can constrain
ourselves to simpler topologies if we are interested in the leading terms in the series.

In particular, the world-sheet topologies in Fig.~\ref{fig:4open} and Fig.~\ref{fig:4closed}
correspond to tree-level diagrams for 4 open and closed strings respectively. Any other topologies
would have holes, therefore, would represent loop-diagrams. Since we want to compare our
calculation here with a tree-level result in QFT, we will be interested in these tree-level
topologies. Note also, that unlike the Feynman diagrams, where we had 3 different ones for
different channels, the one topology in Fig.~\ref{fig:4closed}, for example, will contain all
the 3 possibilities. It follows from the fact that the string interaction is not exactly localized
on the world sheet, and we can get different strings merging into intermediate one by just
streching the world sheet differently, which will still be the same topology. Therefore, all
the possible tree-level interactions of 4 closed strings by a closed string is represented
by one topology - a sphere.

\subsection{String states}
We will describe now the string states that we will need for the calculation.
In string theory all the different
particles are just different excitations of a string, so we need to find what excitations correspond
to our primary particles of interest: gravitons and massless scalars. It turns out, that if
we look for massless bosons, of spin-2 and spin-0, we find that both correspond to massless
closed strings in NS-NS
\footnote{In superstring theory there are two types of string ground states,
therefore, the string states, created by excitations of a ground state, fall into two disctinct
sets, which are called \emph{sectors}. One is called
Neveu-Schwarz (NS) sector and contains bosons. The other one, Ramond (R) sector, contains fermions.
Closed strings are basically composed of two such states and, therefore, there are four sectors:
NS-NS and R-R are bosons, and NS-R and R-NS are fermions.}
sector
with different polarizations. Massless NS-NS
states are constructing by acting on a $k$-momentum NS-NS ground state with two vector creation
operators \cite{pol2}:
\begin{equation}
\psi^\mu_{-1/2}\wt{\psi}^\nu_{-1/2}\ket{0;k}_{\text{NS-NS}},
\end{equation}
A general physical massless closed string state is then described by momentum $k^\mu$ and
polarization 2-tensor $e_{\mu\nu}$:
\begin{equation}
\ket{e;k}_{\text{NS-NS}}=e_{\mu\nu}\psi^\mu_{-1/2}\wt{\psi}^\nu_{-1/2}\ket{0;k}_{\text{NS-NS}},
\label{eq:closed_state}
\end{equation}
which in addition have to satisfy the requirements \cite{pol1}:
\begin{equation}
k^2=k^\mu e_{\mu\nu}=k^\nu e_{\mu\nu}=0,
\label{eq:econstraints}
\end{equation}
and the states are physically identified under:
\begin{equation}
e_{\mu\nu}\cong e_{\mu\nu}+a_\mu k_\nu + k_\mu b_\nu, \quad
a\cdot k=b\cdot k=0.
\end{equation}
The reason why these types of excitations represent more than just one particle, is that the
2-tensor representation $e_{\mu\nu}$ by which the particles transform under the
\emph{little group} \cite{weinberg} $SO(D-2)$ that leaves the momentum invariant, is not
an irreducible representation. It decomposes \cite{pol1,zwiebach}
 into a \emph{symmetric traceless tensor}:
\begin{equation}
e^{(g)}_{\mu\nu}=e^{(g)}_{\nu\mu},\quad \eta^{\mu\nu}e^{(g)}_{\mu\nu}=0,
\label{eq:esym}
\end{equation}
an \emph{antisymmetric tensor}:
\begin{equation}
e^{(b)}_{\mu\nu}=-e^{(b)}_{\nu\mu},
\label{eq:easym}
\end{equation}
and a \emph{trace part}, which is invariant under the $SO(D-2)$ \cite{gs}:
\begin{equation}
e^{(\phi)}_{\mu\nu}=\frac{1}{\sqrt{D-2}}\left(
\eta_{\mu\nu}-k_\mu \xi_\nu-\xi_\mu k_\nu \right), \quad
 k\cdot\xi=1.
\label{eq:edil}
\end{equation}
These three components each satisfy the constraints (\ref{eq:econstraints}) and don't mix under
the transformations of the little group. The symmetric traceless part $e^{(g)}$ corresponds to the
\emph{graviton} and the trace part $e^{(\phi)}$ is $SO(D-2)$ invariant, thus a \emph{massless scalar}
particle, which in string theory is called \emph{dilaton}. Therefore, we will be looking for
an amplitude for 4 $e^{(\phi)}$-polarized closed string scattering by a $e^{(g)}$-polarized closed
string. Note that we chose the normalization for the explicit expression $e^{(\phi)}_{\mu\nu}$ such
that
\begin{equation}
e_{\mu\nu}e^{\mu\nu}=1.
\end{equation}
Such normalization is assumed by the expressions for vertex operators and amplitudes that we will cite.

Even though we don't necessarily have to deal with open strings in this calculation it will be
very convenient to use them
for an intermediate step in the calculation. It is because, as we will see, the amplitude for a closed
string essentially factors into a product of two open-string amplitudes, which are in turn
twice as less complicated. For this calculation we will need to know the states of massless
open strings. Again we are interested in the NS sector, as the amplitude of NS-NS closed strings
will be composed of two NS open string amplitudes. The general state here is \cite{pol2}:
\begin{equation}
\ket{e;k}_{\text{NS}}=e_\mu \psi^{\mu}_{-1/2}\ket{0;k}_{\text{NS}},
\label{eq:open_state}
\end{equation}
with
\begin{equation}
k^2=k^\mu e_\mu=0,
\end{equation}
and identification:
\begin{equation}
e_\mu\cong e_\mu+\gamma k_\mu.
\end{equation}
The states here transform as a vector $e_\mu$, thus are irreducible and represent one particle.
This massless vector boson is actually identified with a photon. Again, it will be assumed
that $e_\mu e^\mu=1$.

\subsection{Open string tree amplitudes}
We proceed now with calculation of the amplitudes for open string scattering - in particular,
we will calculate tree-level amplitudes for interactions of 3 and 4 massless bosons described by the
string state (\ref{eq:open_state}).
We are primarily interested in the \emph{closed} string scattering, however, the results
in this section will allow us to get the closed string amplitudes quickly.

First consider 3 open string interaction. The tree-level world surface for this process
is a disk with three vertex operators at the boundary. This either describes 2 strings
joining or 1 string splitting into two - the amplitude we calculate is general, because
whether the string is incoming or outgoing depends only on the momentum $k$ in the vertex
operator - for outgoing string $k$ is the negative of the actual string momentum.

From the conformally equivalent shapes to represent a disk, for all explicit expressions
we fix it to be the upper half of the complex plane, with the boundary being the real line.
The vertex operators for the open string state (\ref{eq:open_state}) inserted at position
$y\in(-\infty,\infty)$ on the world-sheet boundary (real line) are then \cite{pol2}:
\begin{align}
\label{eq:vo1}
{\cal V}^{-1}(e;k;y) &= g_o e^{-\phi} e_\mu \psi^\mu e^{i k\cdot X}(y),\\
{\cal V}^0 (e;k;y) &= g_o (2\alpha')^{-1/2} e_\mu (i\dot{X}^\mu+2\alpha'k\cdot\psi\psi^\mu)
e^{i k\cdot X}(y),
\label{eq:vo0}
\end{align}
where the fields are all at $y$, and the products of fields at the same point are normal-ordered.
 The two operators are in -1 and 0 ``pictures" - it turns out that when calculating an
amplitude the sum of pictures has to be -2. Note that in addition to $X^\mu$ and $\psi^\mu$ fields
we have a $\phi$ ghost field.
Furthermore, we will have the ${\cal V}$'s multiplied by another ghost field $c$, in
case their positions $y$ on the world sheet are \emph{gauge-fixed}. This gauge-fixing
comes about as follows. In action (\ref{eq:sworlds}) the positions of the vertices are
a property of the world sheet ${\cal W}$, therefore are integrated over, as different
world sheets. However some of the different configurations are related by conformal
transformations and are equivalent. It turns out that by fixing the gauge we can
fix the positions of 3 vertices - these get multiplied by a ghost field $c$. The ones
that are left over (one, in the case of 4 strings) have to have their positions
integrated as part of the moduli ${\cal M}$ of the topology in (\ref{eq:stopologies}).
In addition, different cyclic ordering of the three fixed coordinates on the boundary
of the disk are not actually conformally equivalent, so we have to sum over the
two possibilities also as part of the moduli (discreet, this time). Actually, the
``pictures" and the $\phi$ field come from a similar gauge fixing, but of an
\emph{anticommuting} coordinate \cite{pol2}, that we didn't discuss.

Finally, note the undetermined constant $g_o$ appearing in ${\cal V}$'s, which
should make the states correctly normalized. It can be related later to
other unknown constants, but in the end we are still left with one constant,
that quantifies the strength of open string interactions, and we choose it to be $g_o$.
Also we have another fundamental constant of string theory $\alpha'$.

With all these considerations the tree-level amplitude for 3 open strings looks like:
\begin{align}
\label{eq:so3_1}
S_o^{3}&=\int {\cal D}X{\cal D}\psi{\cal D}\phi{\cal D}c\,
\exp(i S[{\cal W},X,\psi,\phi,c])
c{\cal V}^{-1}_1(y_1) c{\cal V}^{-1}_2(y_2) c{\cal V}^{0}_3(y_3)\\
&+({\cal V}_1\leftrightarrow {\cal V}_2).
\nonumber
\end{align}
The additional term with ${\cal V}_1$ and ${\cal V}_2$ interchanged places the same
strings only in cyclically reversed order on the boundary of the world sheet,
which gives the other, conformally-inequivalent term.
The values for $y_i$ can be chosen arbitrarily and the $S_o^3$ shouldn't depend on that.

We note some facts about the action $S$ to proceed further. We see that it depends
on all the fields and on the world sheet itself. The explicit ${\cal W}$-dependence
of $S$ is just a constant factor having to do with the curvature of the world sheet. For
our cases of interest the value is $e^{-\lambda}$ for a disk and $e^{-2\lambda}$ for
a sphere with some constant $\lambda$. Furthermore, the remaining action splits into
the sum of actions for different fields:
\begin{equation}
S_{\cal W}[X,\psi,\phi,c]=\left(S_X[X]+S_\psi[\psi]+S_\phi[\phi]+S_c[c]\right)_{\cal W},
\end{equation}
so the integral giving the amplitude factors into integrals over different fields. We
note by subscript ${\cal W}$ that the field-action is still implicitly world-sheet
dependent (it matters where the fields are defined), so when we cite some
result, we indicate for what world-sheet it is valid.

To shorten the
notation we will write the integrals as \emph{expectation values}.
By the expectation value $\left<{\cal O}\right>_{\cal W}$ we mean only to include the integral
and the action of the fields that appear in ${\cal O}$, that is:
\begin{equation}
\left<{\cal O}\right>_{\cal W} = \int {\cal D}[\Phi]\, {\cal O} \exp(i S_{\Phi}),
\end{equation}
if $\Phi$ is the set of fields that ${\cal O}$ depends on. The subscript ${\cal W}$ for
the expectation value indicates the world-sheet that the fields live on.

Factoring out the explicit topology-dependent factor, we then write the amplitude
(\ref{eq:so3_1}) as:
\begin{equation}
S_o^3=e^{-\lambda}\left< c{\cal V}^{-1}_1(y_1) c{\cal V}^{-1}_2(y_2)
 c{\cal V}^{0}_3(y_3)\right>_{D}
+({\cal V}_1\leftrightarrow {\cal V}_2),
\label{eq:so3_2}
\end{equation}
where $D$ indicates the disk world-sheet. We will drop this index when it's obvious
what world-sheet we are considering.
Plugging in the vertex operator expressions (\ref{eq:vo1}), (\ref{eq:vo0}) into
(\ref{eq:so3_2}) and factoring some fields out we get:
\begin{align}
\nonumber
S_o^3&=e^{-\lambda}g_o^3(2\alpha')^{-1/2}
\left<c(y_1)c(y_2)c(y_3)\right>
\left<e^{-\phi}(y_1)e^{-\phi}(y_2)\right>\\
\nonumber
&\times e_{1\mu}e_{2\nu}e_{3\rho}
\underbrace{\left<\psi^\mu e^{ik_1\cdot X}(y_1) \psi^\nu e^{ik_2\cdot X}(y_2)
(i\dot{X}^\rho+2\alpha'k_3\cdot\psi\psi^\rho)e^{i k_3\cdot X}(y_3)\right>}
_{\equiv E_X}\\
&+({\cal V}_1\leftrightarrow {\cal V}_2).
\label{eq:so3_3}
\end{align}
Here interchanging the ${\cal V}$'s just means the interchanging $k$'s and $e$'s.
We can further factor $E_X$ which still contains two fields, $X$ and $\psi$:
\begin{align}
\nonumber
E_X^{\mu\nu\rho}&=
i \left< \psi^\mu(y_1)\psi^\nu(y_2)\right>
\left<e^{ik_1\cdot X}(y_1)e^{ik_2\cdot X}(y_2)\dot{X}^\rho e^{ik_3\cdot X}(y_3)\right>\\
&+2\alpha'\left< \psi^\mu(y_1)\psi^\nu(y_2)k_3\cdot\psi\psi^\rho(y_3)\right>
\left<e^{ik_1\cdot X}(y_1)e^{ik_2\cdot X}(y_2)e^{ik_3\cdot X}(y_3)\right>.
\end{align}

Now in order to evaluate $S_o^3$ we need the expectation values for these different
fields on a disk. For fields $X$ and $\psi$ we will use Wick's theorem to sum over
all contractions,
so we will list rules for contracting by writing $\left<\underbracket[0.5pt]{XY}\dots\right>$
to mean the term when contracting $X$ and $Y$. Also note that fields at the same
point on the world-sheet are always normal-ordered, and so are not contracted
with each other (the term gives zero).
Then the values that we need are \cite{pol1,pol2}:
\begin{align}
(\text{notation:}\quad y_{ij}&\equiv y_i-y_j)\\
\label{eq:expectations1}
\left<\prod_i e^{ik_i \cdot X(y_i)}\right>_D&=
iC_D^X(2\pi)^D\delta^D(\Sigma_i k_i)\prod_{i<j}|y_{ij}|^{2\alpha'k_i\cdot k_j},\\
\left<\underbracket[0.5pt]{\dot{X}^\mu(y_1)e}\!^{i k\cdot X(y_2)}\dots\right>_D&=
-2i\alpha'\frac{k^\mu}{y_{12}}\left<e^{i k\cdot X(y_2)}\dots\right>_D,\\
\left<\underbracket[0.5pt]{\dot{X}^\mu(y_1)\dot{X}}\!^\nu(y_2)\dots\right>_D&=
-2\alpha'\frac{\eta^{\mu\nu}}{y_{12}^2}\left<\dots\right>_D,\\
\left<\underbracket[0.5pt]{\psi^\mu(y_1)\psi}\!^\nu(y_2)\dots\right>_D&=
\frac{\eta^{\mu\nu}}{y_{12}}\left<\dots\right>_D,\quad\text{with }\left<1\right>_{\psi,D}=1,\\
\left<c(y_1)c(y_2)c(y_3)\right>_D&=C_D^g |y_{12}y_{13}y_{23}|, \\
\left<e^{-\phi(y_1)}e^{-\phi(y_2)}\right>_D&=\frac{1}{|y_{12}|},
\label{eq:expectations2}
\end{align}
and there exists the relation between the constants \cite{pol1}:
\begin{equation}
C_D\equiv e^{-\lambda}C_D^X C_D^g=\frac{1}{\alpha' g_o^2},
\end{equation}
that will allow us to leave only $g_o$ and $\alpha'$ in the amplitudes. With those
expectation values we can pretty straightforwardly evaluate (\ref{eq:so3_3}),
summing over possible contractions when needed. For example in $E_X$ we have:
\begin{align}
\nonumber
\left<\psi^\mu(y_1)\psi^\nu(y_2)k_3\cdot\psi\psi^\rho(y_3)\right>&=
\left<\underbracket[0.5pt]{\psi^\mu_1\psi}\!^\rho_3
\underbracket[0.5pt]{\psi^\nu_2 k_3\cdot\psi}\!_3\right>-
\left<\underbracket[0.5pt]{\psi^\mu_1 k_3\cdot\psi}\!_3
\underbracket[0.5pt]{\psi^\nu_2\psi}\!^\rho_3\right>\\
&=\frac{\eta^{\mu\rho}}{y_{13}}\frac{k^\nu_3}{y_{23}}-
\frac{k^\mu_3}{y_{13}}\frac{\eta^{\nu\rho}}{y_{23}},
\end{align}
where in the first equality the minus sign comes from permuting anticommuting
fields and we indicate the coordinate by subscript (we will use this notation
for brevity when not ambiguous). An example of how $X$ fields contract in $E_X$
is:
\begin{align}
\nonumber
\left<e^{ik_1\cdot X_1}e^{ik_2\cdot X_2}\dot{X}_3^\rho e^{ik_3\cdot X_3}\right>&=
\left<\underbracket[0.5pt]{e^{ik_1\cdot X_1}e^{ik_2\cdot X_2}\dot{X}}\!_3^\rho e^{ik_3\cdot X_3}\right>+
\left<e^{ik_1\cdot X_1}\underbracket[0.5pt]{e^{ik_2\cdot X_2}\dot{X}}\!_3^\rho e^{ik_3\cdot X_3}\right>\\
&=-2i\alpha'\left(\frac{k_1^\rho}{y_{31}}+\frac{k^\rho_2}{y_{32}}\right)
\left<e^{ik_1\cdot X_1}e^{ik_2\cdot X_2}e^{ik_3\cdot X_3}\right>\\
&=2C_D^X\alpha'(2\pi)^D\delta^D(\Sigma_i k_i) \left(\frac{k_1^\rho}{y_{31}}+\frac{k^\rho_2}{y_{32}}\right)
\prod_{i<j}|y_{ij}|^{2\alpha'k_i\cdot k_j}.
\end{align}
Note an important identity for 3 massless string scattering:
\begin{equation}
k_i\cdot k_j=0=\frac{1}{2}\left((k_i+k_j)^2-k_i^2-k_j^2\right),
\end{equation}
since the sum of two momenta is minus the third one because of the delta function
(momentum conservation), and $k_i^2=0$ for all three. Another identity, that comes from
$e_i\cdot k_i=0$ (no sum) and $\Sigma_i k_i=0$ is:
\begin{equation}
e_1\cdot k_2=-e_1\cdot k_3,
\end{equation}
and similar for other $e_i$'s.

We can continue to evaluate parts of $S_o^3$ using the above identities for simplification.
In the end the $y_i$'s drop out, as they should
(or we can fix their values from the beginning to simplify the calculation), and
the result is \cite{pol2}:
\begin{align}
S_o^3&=i\frac{g_o}{\sqrt{2\alpha'}}(2\pi)^D\delta^D(\Sigma_i k_i)
\label{eq:so3_4}
e_{1\mu}e_{2\nu}e_{3\rho}V_{123}^{\mu\nu\rho}+({\cal V}_1\leftrightarrow{\cal V}_2),\\
V^{\mu\nu\rho}_{123}&\equiv 2(\eta^{\mu\nu}k_1^{\rho}+\eta^{\nu\rho}k_2^\mu+\eta^{\rho\mu}k_3^\nu).
\label{eq:vdef}
\end{align}
It actually turns out that the interchanged term exactly cancels the first one so $S_o^3=0$
(which is reasonable for the coupling of three photons). However, for closed string calculation
the result without the interchange will be useful. For that we rewrite (\ref{eq:so3_4}) as:
\begin{align}
\label{eq:so3_5}
S_o^3&=(2\pi)^D\delta^D(\Sigma_i k_i)[A_o^3(1;2;3)+A_o^3(2;1;3)],\\
A_o^3(a;b;c)&=i\frac{g_o}{\sqrt{2\alpha'}}e_{a\mu}e_{b\nu}e_{c\rho}V_{abc}^{\mu\nu\rho},
\label{eq:ao3}
\end{align}
and later we will use $A_o^3$. Note that the $A$-amplitude, which we get from the S-matrix
$S$ by factoring out $2\pi$'s and $\delta$, is exactly the analog of $i{\cal M}$ in field
theory (\ref{eq:smatrix}). Since in field theory the vertex amplitude is the $i{\cal M}$
for the process described by that vertex, we can directly compare amplitude $A^3$ from
string theory with the corresponding field theory vertex. We will be able to do that
for 3-closed string scattering.

Now lets move to 4-open strings. All the discussion above applies directly here, so now
it is just a direct calculation. In terms of vertex operators it is:
\begin{equation}
S_o^4=e^{-\lambda}\int d y_4
\left<c{\cal V}_1^{-1}(y_1)c{\cal V}_2^{-1}(y_2)c{\cal V}_3^0(y_3){\cal V}_4^0(y_4)\right>+
({\cal V}_1\leftrightarrow{\cal V}_2),
\end{equation}
note that one position $y_4$ is not fixed, and there is no corresponding $c$-field. We
can straightforwardly expand the expression and evaluate the expectation values. For the
evaluation of the integral it is convenient to fix values $y_1=0$, $y_2=1$,
$y_3\rightarrow\infty$. As for
QFT 4-massless scattering, the momenta here satisfy $k_i^2=0$ and $\Sigma_i k_i=0$ (we
get the $\delta$-function again from expectation of exponentials), therefore, it is
useful again to use Mandelstam variables (\ref{eq:2.46}).
In the end we get the result \cite{pol2}:
\begin{align}
\label{eq:342}x
S_o^4=&-8ig_o^2\alpha'(2\pi)^D\delta^D(\Sigma_i k_i)K_o F_o,\\
\nonumber
K_o\equiv&-\frac14 (ute_1\cdot e_2\, e_3\cdot e_4+(2\text{ perm.}))\\
&-\frac12 (se_3\cdot k_2\, e_1\cdot k_4\, e_2\cdot e_4+(11\text{ perm.}))
\label{eq:ko}
\equiv\,e_{1\mu}e_{2\nu}e_{3\rho}e_{4\sigma}K^{\mu\nu\rho\sigma},\\
F_o\equiv&\frac{\Gamma(-\alpha' s)\Gamma(-\alpha' t)}{\Gamma(1+\alpha' u)}+
\frac{\Gamma(-\alpha' t)\Gamma(-\alpha' u)}{\Gamma(1+\alpha' s)}+
\frac{\Gamma(-\alpha' u)\Gamma(-\alpha' s)}{\Gamma(1+\alpha' t)}.
\label{eq:344}
\end{align}
The permutations in the expression for $K_o$ mean that we should add all \emph{inequivalent}
terms that we can get from the first one by permuting indices $\{1,2,3,4\}$ on $k_i$ and $e_i$
simultaneously.
For the first term the 2 others we get by $(2\leftrightarrow 3)$ and $(2\leftrightarrow 4)$.
The second term is only identical under $((1\leftrightarrow 3),(2\leftrightarrow 4))$
interchange, so we have a total of $24/2=12$ terms, where 24 is the size of 4-permutation
group. The $\Gamma$'s in $F_o$ appear from $y_4$ integration, and they contain all the relevant
poles associated with the intermediate string - we will examine such factor in more detail for the
closed string. Finally, note that $K^{\mu\nu\rho\sigma}$ is implicitly defined in (\ref{eq:ko})
as the factor multiplying the corresponding components of $e_i$'s. This definition will be used
for the closed string.

\subsection{Closed string tree amplitudes}
We proceed now to our final goal - calculation of 4 dilaton scattering. We will have
to do it in a couple of steps. First, we will calculate the general amplitude for 4
massless closed string scattering by another closed string. Then, by choosing the
dilaton polarizations for the 4 strings we will have 4 dilaton scattering amplitude
by \emph{arbitrary} closed string. This will include more than just graviton. We will
then have to analyze the amplitude in some detail to pick out the part due to the graviton.

The amplitude calculation for closed string is analogous to the open string in previous
section. One difference is that this time the world-sheet is a sphere, as in
Fig.~\ref{fig:4closed}, which will introduce some changes in expectation values.
For explicit expressions we fix the sphere to be represented by the full complex plane
with coordinate $z$. First, we need the
vertex operators that for the massless closed string states (\ref{eq:closed_state}) are
\cite{pol2}:
\begin{align}
\label{eq:vc1}
{\cal V}^{-1,-1}(z)&=g_c e^{-\phi-\wt{\phi}}e_{\mu\nu}\psi^\mu\wt{\psi}^\nu e^{ik\cdot X}(z),\\
{\cal V}^{0,0}(z)&=-\frac{2 g_c}{\alpha'}e_{\mu\nu}
(i\partial_z X^\mu+\frac{\alpha'}2 k\cdot\psi\psi^\mu)
(i\partial_{\bar{z}} X^\nu+\frac{\alpha'}2 k\cdot\wt{\psi}\wt{\psi}^\nu)e^{ik\cdot X}(z),
\end{align}
where (-1,-1) and (0,0) again are different pictures and we want the total sum in the
amplitude to be (-2,-2). Note the different constant of normalization $g_c$ which will
in the end quantify the closed string coupling strength. As with the open strings,
3 coordinates of vertex operators will be gauge-fixed in an amplitude and those operators
will be multiplied by ghost fields $c\wt{c}$. The other coordinates will be integrated
over the complex plane. Note that unlike for the disk, where there were two inequivalent
cyclic orderings of fixed coordinates, on a sphere any positioning of 3 points is equivalent
 so there will be only one term.

We can see already that the operators are analogous to the open strings, except that
there are two sets of fields $\{\psi,\phi,c\}$ and $\{\wt{\psi},\wt{\phi},\wt{c}\}$,
which are called left-moving and right-moving respectively. Their actions are independent
so the expectation values factor. There is only one field $X$, but the derivatives
$\partial_z X^\mu$ and $\partial_{\bar{z}} X^\mu$ are again independent in a sense
that the expectation value between them is zero, so the expectation values for $X$
will also factor into left- and right-moving parts. It is true, though, that there is only
one set of $e^{ik\cdot X}$, that will give an overall factor to the amplitude.

We won't list the expectation values for these fields on a sphere, but they are essentially
the same as (\ref{eq:expectations1})-(\ref{eq:expectations2}) for the left- and right-moving
parts separately, with $y$ replaced by $z$ for left- and with $\bar{z}$ for right-moving,
and also with $\alpha'\rightarrow \alpha'/4$.  The corresponding constants are also
replaced by new ones: $C_S^X$ and $C_S^g$ with a relation \cite{pol1}:
\begin{equation}
C_S\equiv e^{-2\lambda}C_S^X C_S^g=\frac{8\pi}{\alpha' g_c^2}.
\end{equation}

With all this in mind it is clear that the expectation value for closed string
vertex operators \emph{before integration} is two copies of open expectation values:
one from left-moving and one from right-moving parts \cite{gs,pol1,kawai}. The difference
is only in
constants and in that there is only one common $\delta$-function. The integration,
in case of 4 strings, can give more non-trivial factors.

Taking care of the constant factors we get the following relations between
$A$-amplitudes for 3 strings:
\begin{equation}
A^3_c=-i\pi\frac{\alpha'}{2}\frac{g_c}{g_o^2}
\left[A_o^3\left(\frac{\alpha'}4\right)\otimes\wt{A}^3_o\left(\frac{\alpha'}4\right)\right],
\end{equation}
and for 4 strings \cite{pol2,gs,kawai}:
\begin{equation}
A^4_c=i\pi^2\alpha'\frac{g_c^2}{g_o^4}
\frac{1}{\Gamma\left(-\frac{\alpha'}{4}t\right)\Gamma\left(1+\frac{\alpha'}{4}t\right)}
\left[A_o^4\left(s,t;\frac{\alpha'}4\right)\otimes\wt{A}^4_o\left(t,u;\frac{\alpha'}4\right)\right].
\end{equation}
By $\otimes$ product we mean that the polarizations from the two sides combine as
\begin{equation}
e_{1\mu}\otimes \wt{e}_{1\nu}=e_{1\mu\nu},
\end{equation}
because in the vertex operators (\ref{eq:vc1}) it is
the overall polarization multiplying both parts. Then by $\wt{A}$ is just meant that
it contains $\wt{e}$'s - the ``right side" of polarizations.
 For 3 strings $A_o^3(\alpha'/4)$ is just
$A^3_o(1;2;3)$ in (\ref{eq:ao3}) with replaced $\alpha'\rightarrow\alpha'/4$. For 4 string
amplitudes we are supposed to take as $A_o^4$ the contribution to (\ref{eq:342})
from one of the 6 possible
cyclic ordering of strings on the boundary. Different orderings give different terms in $F_o$,
with each term in (\ref{eq:344}) being the sum of two identical contributions.
The argument in $A_o^4$ indicates which poles we should choose, so then:
\begin{align}
A_o^4\left(s,t;\frac{\alpha'}4\right)&=
-ig_o^2\alpha'K_o
\frac{\Gamma(-\frac{\alpha'}4 s)\Gamma(-\frac{\alpha'}4 t)}{\Gamma(1+\frac{\alpha'}4 u)},\\
\wt{A}_o^4\left(t,u;\frac{\alpha'}4\right)&=
-ig_o^2\alpha'\wt{K}_o
\frac{\Gamma(-\frac{\alpha'}4 t)\Gamma(-\frac{\alpha'}4 u)}{\Gamma(1+\frac{\alpha'}4 s)},
\end{align}
while $S_o^4$ can be expressed as:
\begin{equation}
S_o^4=(2\pi)^D\delta^D(\Sigma_i k_i)
\left(2 A_o^4(s,t;\alpha')+2 A_o^4(t,u;\alpha')+2 A_o^4(u,s;\alpha')\right).
\end{equation}

Using now these relations between open and closed amplitudes we can write directly the
amplitudes for 3 closed and 4 closed strings \cite{pol2}:
\begin{align}
\label{eq:ac3}
A^3_c&=i\pi g_c V_c,\\
\label{eq:vc}
V_c&\equiv e_{1\mu_1\nu_1}e_{2\mu_2\nu_2}e_{3\mu_3\nu_3}V^{\mu_1\mu_2\mu_3}V^{\mu_1\mu_2\mu_3},\\
A^4_c&=-i\pi^2 g_c^2 {\alpha'}^3 K_c F_c,\\
\label{eq:kc}
K_c&=e_{1\mu_1\nu_1}e_{2\mu_2\nu_2}e_{3\mu_3\nu_3}e_{4\mu_4\nu_4}
K^{\mu_1\mu_2\mu_3\mu_4}K^{\nu_1\nu_2\nu_3\nu_4},\\
F_c&=\frac{\Gamma(-\frac{\alpha'}{4}s)\Gamma(-\frac{\alpha'}{4}t)\Gamma(-\frac{\alpha'}{4}u)}
{\Gamma(1+\frac{\alpha'}{4}s)\Gamma(1+\frac{\alpha'}{4}t)\Gamma(1+\frac{\alpha'}{4}u)},
\end{align}
with $V^{\mu\nu\rho}$ defined in (\ref{eq:vdef}) and $K^{\mu\nu\rho\sigma}$ in (\ref{eq:ko}).
The S-matrix amplitude for closed strings is related for both 3 and 4 as:
\begin{equation}
S_c^{3,4}=(2\pi)^D\delta^D(\Sigma_i k_i) A_c^{3,4}.
\end{equation}
There is no sum over interchanges as for open strings (\ref{eq:so3_5}), so $A_c$ gives the whole
tree-level term corresponding to $i{\cal M}$ in QFT.

\subsection{Dilaton amplitude}
Now that we have a general expression for the 4 closed string scattering amplitude, we can
get the one we are primarily interested in - 4 dilaton interaction. We simply have to plug in
the polarization (\ref{eq:edil}) in the expression (\ref{eq:kc}) for $K_c$. The calculation
is rather tedious, note that we have two copies of  $K^{\mu\nu\rho\sigma}$, which is given by
permutations in (\ref{eq:ko}), contracted together by $e_{\mu\nu}^{(\phi)}$'s:
\begin{align}
K_c^{(\phi)}&=\frac{1}{(D-2)^2}(\eta_{\mu_1\nu_1}-k_{1\mu_1} \xi_{1\nu_1}-\xi_{1\mu_1} k_{1\nu_1})
(\eta_{\mu_2\nu_2}-k_{2\mu_2} \xi_{2\nu_2}-\xi_{2\mu_2} k_{2\nu_2})\\
&\times(\eta_{\mu_3\nu_3}-k_{3\mu_3} \xi_{3\nu_3}-\xi_{3\mu_3} k_{3\nu_3})
(\eta_{\mu_4\nu_4}-k_{4\mu_4} \xi_{4\nu_4}-\xi_{4\mu_4} k_{4\nu_4})
K^{\mu_1\mu_2\mu_3\mu_4}K^{\nu_1\nu_2\nu_3\nu_4},
\nonumber
\end{align}
but again using Mandelstam variables and the momentum conservation the
result in the end simplifies to:
\begin{equation}
K_c^{(\phi)}=\frac{1}{16}(t^2u^2+u^2s^2+s^2t^2).
\end{equation}
We can then put together the total tree-level amplitude for 4 dilaton scattering by
\emph{any} closed string:
\begin{equation}
A_c^{4(\phi)}=-\frac{i\pi^2 g_c^2 {\alpha'}^3}{16}
\frac{\Gamma(-\frac{\alpha'}{4}s)\Gamma(-\frac{\alpha'}{4}t)\Gamma(-\frac{\alpha'}{4}u)}
{\Gamma(1+\frac{\alpha'}{4}s)\Gamma(1+\frac{\alpha'}{4}t)\Gamma(1+\frac{\alpha'}{4}u)}
(t^2u^2+u^2s^2+s^2t^2).
\label{eq:af_1}
\end{equation}
Again, we emphasize that this amplitude contains not only gravitational interaction. We can
get insight into what particles mediate the interaction by looking at the poles of the
amplitude - they appear when the intermediate particle is on-shell, therefore, we can
deduce the masses of the particles that contribute to the interaction.

This information is contained in the function $F_c$ (note that it is polarization-independent
so this analysis of poles is valid for any 4-closed string interaction). Function
$\Gamma$ has poles at:
\begin{equation}
|\Gamma(x)|\rightarrow\infty\ \Leftrightarrow\
x=-n, \quad n=0,1,2,\dots,
\end{equation}
and $\Gamma(x)$ is never 0. So then $F_c(s,t,u)$ will have poles:
\begin{equation}
|F_c(s,t,u)|\rightarrow\infty\ \Leftrightarrow\
s,t,u=\frac{4}{\alpha'}n, \quad n=0,1,2,\dots,
\end{equation}
where we mean that \emph{any} of the three variables satisfies the condition.
These are also the poles of the whole amplitude\footnote{Note that
the $K_c$ term in general does not cancel the pole, because if, let's say, $s=0$ then
$t=-u$ and $K_c^{(\phi)}\propto t^4$. This is not 0 unless $s=t=u=0$, which is a special case
that we ignore (it has to be taken as a limit of our amplitude).}.
As mentioned in the discussion of the QFT amplitude, $s,t,u=m^2$ - the mass of the
intermediate particle in each of the channels. Since $A_c^4$ takes into account all the
3 channels, it is natural that there are identical poles for each of them, and we can
conclude that the particles mediating the interaction have masses:
\begin{equation}
m^2=\frac{4}{\alpha'}n, \quad n=0,1,2,\dots
\label{eq:mstring}
\end{equation}
This is exactly the spectrum of closed strings in the theory \cite{pol2}. Appart being a check
that the result is consistent, this tells us that the massless closed strings couple to closed
strings of any mass.

Now we want to limit the intermediate particle to being a graviton. The first thing we can
do is limiting it to be \emph{massless} - that could be in general a graviton, an antisymmetric
ensor, or another dilaton.
This is done easily by taking the low-energy limit of the string
theory by letting $\alpha'\rightarrow 0$. Since the masses of strings are proportional to
$1/\alpha'$, as in (\ref{eq:mstring}) and similarly for open strings, this limit only allows
massless strings to be created, which is what we want. Expanding $F_c$ in series of $\alpha'$
gives \cite{gs,pol2}:
\begin{equation}
F_c=-\frac{64}{{\alpha'}^3stu}+O({\alpha'}^0).
\end{equation}
Plugging this back in (\ref{eq:af_1}) gives:
\begin{equation}
A_c^{4(\phi)}=4i\pi^2 g_c^2
\left(\frac{tu}{s}+\frac{us}{t}+\frac{st}{u}\right)+O({\alpha'}^3),
\label{eq:af_2}
\end{equation}
so for $\alpha'\rightarrow 0$ we simply get this first, $\alpha'$-independent term.
This tells us the amplitude for dilaton scattering by a \emph{massless} closed
string. We can already see the correspondence with QFT amplitude (\ref{eq:sqft}),
and, actually, this $A_c^{4(\phi)}$ \emph{will} turn out to be just the gravitational
amplitude, but we still have to show that.

 What we want is to analyze the coupling of two dilatons with a third massless
closed string, which we can do by looking at the 3-string amplitude (\ref{eq:ac3})
with appropriate polarizations. Such process is a part of 4-string amplitude in
any of the channels (consider the Feynman diagrams from QFT). If we find that
the scattering amplitude of two dilatons into a third particle of some kind is 0,
it means that this intermediate particle can not contribute to $A_c^{4(\phi)}$.

We calculate then the amplitude $A_c^3$ with two of the polarizations taken to be
dilatons and the third one - arbitrary $e_{\mu\nu}$.
Plugging the values into $V_c$ in (\ref{eq:vc}) we have:
\begin{align}
V_c^{(\phi)}&=\frac{4}{D-2} e_{\mu_1\nu_1}
(\eta_{\mu_2\nu_2}-k_{2\mu_2} \xi_{2\nu_2}-\xi_{2\mu_2} k_{2\nu_2})
(\eta_{\mu_3\nu_3}-k_{3\mu_3} \xi_{3\nu_3}-\xi_{3\mu_3} k_{3\nu_3})\\
&\times
(\eta^{\mu_1\mu_2}k_1^{\mu_3}+\eta^{\mu_2\mu_3}k_2^{\mu_1}+\eta^{\mu_3\mu_1}k_3^{\mu_2})
(\eta^{\nu_1\nu_2}k_1^{\nu_3}+\eta^{\nu_2\nu_3}k_2^{\nu_1}+\eta^{\nu_3\nu_1}k_3^{\nu_2}).
\end{align}
Using 3-massless particle relations $k_i\cdot k_j=e_{i\mu\nu} k_i^\mu=e_{i\mu\nu} k_i^\nu=0$
and momentum conservation this evaluates to:
\begin{equation}
V_c^{(\phi)}=-2e_{\mu\nu}(k_2^\mu k_3^\nu+k_3^\mu k_2^\nu).
\label{eq:vcf}
\end{equation}
Now we can easily see what happens with different $e_{\mu\nu}$'s. The tensor multiplying
polarization is symmetric so for antisymmetric $e^{(b)}_{\mu\nu}$ the amplitude is immediatly 0.
For dilaton polarization $e^{(\phi)}_{\mu\nu}$ we only get dot-products of $k_i$'s in (\ref{eq:vcf}),
so it is also zero. Therefore, two dilatons couple only to a graviton, among massless
states,
in which case the amplitude is:
\begin{equation}
A_c^{3(\phi)}=-2\pi i g_c e^{(g)}_{\mu\nu}(k_2^\mu k_3^\nu+k_3^\mu k_2^\nu).
\label{eq:af3}
\end{equation}
This result also proves that the (\ref{eq:af_2}) gives exactly the gravitational scattering
amplitude, which we write as the S-matrix for our final result:
\begin{equation}
S_{\text{grav}}^{4(\phi)}=i(2\pi g_c)^2 (2\pi)^D \delta^D(\Sigma_i k_i)
\left(\frac{tu}{s}+\frac{us}{t}+\frac{st}{u}\right).
\end{equation}
We confirm that it is equal to $S$ calculated in QFT in (\ref{eq:sqft}) with identification
of constants \cite{pol2}:
\begin{equation}
\kappa=2\pi g_c.
\end{equation}
In addition we get a second check: if we write the QFT vertex for 2-dilaton-graviton coupling
$V^{\mu\nu}_{(\phi\phi h)}$ in (\ref{eq:vertex}) as an on-shell scattering amplitude:
\begin{equation}
i{\cal M}=h_{\mu\nu}V^{\mu\nu}_{(\phi\phi h)}=-i\kappa h_{\mu\nu}
(k_1^\mu k_2^\nu+k_2^\mu k_1^\nu),
\end{equation}
the result again matches SST amplitude $A_c^{3(\phi)}$ in (\ref{eq:af3}) with the same
identification $\kappa=2\pi g_c$.

\section{Conclusions}
This finishes our comparison of the quantum field theory and superstring theory amplitudes. We
did get a matching result in SST for the process corresponding to gravitational scattering of
4 massless scalars in QFT, and we found a relation between the constants in the two theories. As
promised in the introduction this does show that we can describe gravitation by SST, therefore,
this theory, being consistent as opposed to QFT of gravity, has a chance of being the unifying
theory describing all particles and interactions.

Looking more generally, however, we only demonstrated this feature for a very specific case.
It is possible to argue much more generally \cite{gs,pol2}, what kind of effective actions
the string theory reduces to in low-energy limit, and what those actions correspond to in
QFT point of view. In that respect, the calculation in this paper is not so much valuable for
the result itself, but more as an exercise, allowing us to go through many important topics in
both quantum field theory and superstring theory, and demonstrating (though, a small part of)
relationships between them.

\subsection*{Acknowledgments}
I am grateful to Professor L\"ust, who agreed to supervise me for this work and the relevant
studies, and who pointed this exercise problem to me. I owe many thanks to Professor Zwiebach who first
introduced me to the string theory at his course at M.I.T., and encouraged me for
further studies in this field.

I learned the most of the material directly relevant to this paper from: \cite{gr} - general relativity;
\cite{ps} - quantum field theory; \cite{pol1,pol2,zwiebach} - string theory.

I thank Max-Planck-Institut f\"ur Physik, that supported the internship, during which the work
was done.

%

\end{document}